\documentclass{elsart}
\usepackage{epsfig}
\usepackage{subfigure}

\newcommand{\PB}{\mbox{PbF$_2$}}

\begin{document}

\begin{frontmatter}
  
\title{Measurements and Simulations of Cherenkov Light in Lead Fluoride
  Crystals\thanksref{diss}}
  \thanks[diss]{This work is part of the doctoral thesis of P. Achenbach.}
  \author[mainz]{P. Achenbach\thanksref{author}},
  \author[mainz]{S. Baunack},
  \author[mainz]{K. Grimm},
  \author[mainz]{T. Hammel},
  \author[mainz]{D. von Harrach},
  \author[mainz]{A. Lopes Ginja},
  \author[mainz]{F.E. Maas},
  \author[mainz]{E. Schilling},
  \author[julich]{H. Str\"oher}

  \address[mainz]{Institut f\"ur Kernphysik, Johannes
  Gutenberg-Universit\"at, Becherweg 45, 55099 Mainz, Germany}
  \address[julich]{Institut f\"ur Kernphysik, Forschungszentrum
  J\"ulich GmbH, 52425 J\"ulich, Germany}
  \thanks[author]{Corresponding author. Tel.: +49 6131 39 22958; fax
  +49 6131 39 22964; e-mail: patrick@kph.uni-mainz.de.}

\begin{abstract}
  The anticipated use of more than one thousand lead fluoride (\PB)
  crystals as a fast and compact Cherenkov calorimeter material in a
  parity violation experiment at MAMI stimulated the investigation of
  the light yield (L.Y.) of these crystals. The number of
  photoelectrons ($p.e.$) per MeV deposited energy has been determined
  with a hybrid photomultiplier tube (HPMT). In response to
  radioactive sources a L.Y.\ between 1.7 and 1.9~$p.e.$/MeV was
  measured with 4\% statistical and 5\% systematic error. The L.Y.\
  optimization with appropriate wrappings and couplings was
  investigated by means of the HPMT. Furthermore, a fast Monte Carlo
  simulation based on the GEANT code was employed to calculate the
  characteristics of Cherenkov light in the \PB\ crystals. The
  computing time was reduced by a factor of 50 compared to the regular
  photon tracking method by implementing detection probabilities as a
  three-dimensional look-up table. For a single crystal a L.Y.\ of
  $2.1~p.e.$/MeV was calculated. The corresponding detector response
  to electrons between 10 and 1000~MeV was highly linear with a
  variation smaller than 1\%.
\end{abstract}
  
\begin{keyword}
   Cherenkov counters; lead fluoride; light yield; photoelectron 
   distributions; Monte Carlo simulations

  {\small {\em PACS classification:} 24.10 -- Lx; 29.40 -- Ka}
\end{keyword}

\end{frontmatter}

\section{Introduction}

The A4 collaboration is preparing a measurement of the parity
violating asymmetry $A_\mathrm{0}$ in elastic scattering of right and
left handed electrons on an unpolarized proton target at the Mainz
Microtron MAMI~[1]. An electromagnetic calorimeter was built in the
years 1999 and 2000 to carry out the precise measurement with a total
accuracy of $\delta A_\mathrm{0} < 5\%$.  In order to obtain a clear
separation of events in the low energy tail of the elastic peak to a
background of photons from $\pi^0$ decays, soft electrons and pions an
energy resolution of $\Delta E/E \le 3.5\% /\sqrt{E[ \mathrm{GeV}]}$
in arrays of $3 \times 3$ detectors and a fast online calibration are
needed. This energy resolution not only depends on intrinsic shower
and leakage fluctuations but also on the effective light yield (L.Y.).
Therefore the number of photoelectrons ($p.e.$) per MeV deposited
energy at a given quantum efficiency of a photocathode is of great
interest. Additionally, the effect of non-linearities in the response,
which might produce a degradation of the energy resolution, should be
taken into account.

In the early nineties lead fluoride in its cubic lattice form
($\beta$-\PB) was discovered as a Cherenkov radiator for
electromagnetic calorimetry~[2--4] because of its high transparency
and compactness. The optical transmittance of the $\beta$-\PB\
crystals extends below 270~nm, their radiation resistance is
moderate~[5] and the A4 collaboration decided to use them in order to
exploit their excellent time response ($< 20$~ns). Since the L.Y.\ of
$\beta$-\PB\ is low compared to scintillating crystals a search for
scintillation in doped and orthorhombic \PB\ was performed in recent
years~[6,7]. The measurements presented in this paper have proved that
the L.Y.\ of good quality crystals is sufficient for their application
in medium and high energy physics experiments.  Two methods of
accessing the L.Y.\ will be presented: on the one hand the use of low
energy radioactive sources to measure photoelectron distributions and
on the other hand a Monte Carlo code to simulate the Cherenkov photon
production and detection in the crystals.

One has to consider that \PB\ only emits few photons per MeV of
deposited energy, which complicates conventional laboratory
measurements. The detection of single photons is usually performed by
means of regular photomultiplier tubes (PMTs) as well as by
ultraviolet-sensitive multi-wire proportional chambers (MWPCs). Both
types of photon sensors are limited in their intrinsic resolution by
fluctuations in the number of secondary electrons produced at the
first dynode of a PMT or in the avalanche around the anode wire of a
MWPC.  This limitation favors the use of the recently reinvented
hybrid photomultiplier tube (HPMT) with its excellent multiple photon
separation and high efficiency. A HPMT consists of a reversely biased
silicon P-I-N diode, in which highly accelerated photoelectrons create
a few thousand electron-hole pairs with much smaller statistical
fluctuations. In Section~2 of this paper it will be shown that a HPMT
allows to study the effective L.Y.\ and related properties of \PB\
crystals.

To realize the second method, the GEANT~3.21 Monte Carlo code~[8] was
used. The transport of Cherenkov photons from the location of their
production to the photocathode requires a step by step tracking
through uniform attenuating media to the nearest boundaries. Usually,
large computing times are required because photons are refracted or
reflected and the tracking has to continue until either absorption,
detection or escape of the photons will occur. In Section~3 of this
paper, a method is described that avoids the tracking as soon as a
look-up table is generated, which tabulates the detection
probabilities of a photon depending on its wavelength, its angle
relative to the primary particle's direction and its longitudinal
location of production.

Section 4 provides the summary of the L.Y.\ measurements and
simulations.

\section{Measurements of the Light Yield}

\subsection{Experimental Arrangements}

For the laboratory measurements at Mainz an electrostatically focused
HPMT manufactured by DEP\footnote{Delft Electronic Products BV, Roden,
The Netherlands} with a photocathode of 19~mm useful input diameter
was employed. The S20 photocathode featured a high quantum efficiency
in the ultraviolet region with 27\% at 270~nm and 25\% at 400~nm. The
HPMT was operated at $-15$~kV accelerating voltage and with an applied
reversed-bias voltage of $+90$~V. Two electrodes at a potential of
$-11$~kV provided the focusing of the released photoelectrons onto a
silicon P-I-N diode. Each bombarding photoelectron led to the creation
of $\approx 3500$ electron-hole-pairs, causing an electrical current
in the diode. This output signal was subsequently amplified by a
low-noise charge pre-amplifier, which was built in the HPMT housing to
avoid any unnecessary cabling capacitances by a direct charge
coupling. Coincidences between the HPMT signal and a scintillating
counter were used to detect the cascading $\gamma$-decay of
\nuc{60}{Co} isotopes. The number of $p.e.$ released from the HPMT
photocathode on a quartz window were counted to obtain the L.Y.\ of
the crystal.  The block diagramme in Fig.~1~(a) illustrates the
geometry of the detectors as well as the applied electronic
components. The \PB\ crystal was read out by the HPMT whereas the
plastic scintillator opposite to the crystal was read out by a regular
PMT. The source was placed between the two detectors.  Most of the
measurements were done with the \nuc{60}{Co} source, but during parts
of the data taking a \nuc{90}{Sr} source was used. Then the
scintillator was read out by two coincidence detectors to define the
trigger, see Fig.~1~(b). This set-up allowed to measure events in
which no particle has hit the \PB\ crystal. A blue LED was used in
separate calibration measurements to obtain the HPMT
characteristics. All detectors were located in a light-tight box. The
crystal and the plastic scintillator were coupled with silicone rubber
pads of Elastosil RT~601 with a high transparency and good coupling
reproducibility. The pre-amplifier signals have been shaped by a
commercially available spectroscopy equipment. The energy spectra were
recorded by Constant Fraction Discriminators (CFDs) and
Analog-to-Digital Converters (ADCs) accessed by a CAMAC bus. The
coincidence measurements exhibited almost no background, but dark
counts contributed to false coincidences, i.e. random signals fell
accidentally into the measurement gate. The dominant contribution to
those counts is the thermal emission of electrons off the
photocathode. Its rate depends on the temperature and the
stabilization time of the HPMT. Thirty minutes after switching the
high voltage on the rate of dark pulses had decreased exponentially to
5\% of the initial rate to about 75 counts per second. After some
hours of stabilization the rate had further decreased to about 50
counts per second.

\subsection{Performance of the HPMT}

Measured photoelectron spectra such as the one shown in Fig.~2 have
been obtained with very short and highly attenuated LED pulses. The
photoelectron distributions are composed of Gaussian shaped peaks
corresponding to the overlap of an integer number of released
photoelectrons and a continuum. More than 10 photoelectron peaks are
clearly separated. A calibration of the obtained spectra in numbers of
photoelectrons versus channels of the ADC was possible with an
accuracy of 0.026\%, since the peak positions showed a very high
degree of linearity.  The contrast function $f=\mathrm{(Peak - Valley)
/ (Peak + Valley)}$ calculated from the photoelectron peaks of the
shown photoelectron distribution is presented in Fig.~3. The straight
solid line at 0.03, which is commonly defined as the limit of peak
resolution, crosses the exponentially fitted data points at 14 $p.e.$,
demonstrating the excellent resolving power of the HPMT. Photon
counting measurements with HPMTs of up to fifteen resolved
photoelectron peaks have been reported by C.~d'Ambrosio~[9]. In
contrast, PMTs could resolve only two or three photoelectron peaks.

The photoelectron peaks have been fitted and single photoelectron
resolutions $\sigma_{\mathrm{meas}} = 10.7$\% ($\approx$ 1.60~keV)
were found almost independent on the peak position. Two effects
contribute to this variance: the fluctuations in the number of
electron-hole pairs $\sigma_{\mathrm{diode}}$ and the electronic noise
$\sigma_{\mathrm{noise}}$. The latter must not be neglected in HPMT
measurements, because the total gain of a HPMT is of the order of a
few thousand whereas the PMT gain is usually of the order of
$10^6-10^7$. The set-up with the strontium source and the plastic
scintillator as a trigger allowed to measure the width of the
pedestals in the ADC distributions of the HPMT spectra. The obtained
variance of 9.5\% ($\approx$ 1.43~keV) of the pedestal peak
corresponds to the electronic noise $\sigma_{\mathrm{noise}}$ and can
be subtracted in quadrature from the measured variance, resulting in
the intrinsic resolution $\sigma_{ \mathrm{diode}} = 0.75$~keV. The
continuum in the spectrum is explained by backscattering of the
accelerated photoelectrons off the diode surface which re-enter the
diode at a smaller angle or with lower energy~[9]. The ohmic contact
being responsible for the backscattering effect is ion implanted and
its thickness amounts to only $0.05~\mu$m. The fraction of
backscattered events could be estimated by calculating the ratio of
the continuous area to the peak area, which was about 80\%.

\subsection{Analysis of the Photoelectron Distributions}

In the first part of this study, several small \PB\ samples of the
dimensions $25 \times 25 \times 25$~mm$^3$ were used. The samples have
been polished on all faces by the manufacturer SICCAS. The average
number of $p.e.$ detected by the HPMT was determined from the
photoelectron distributions using the expression:
\begin{displaymath}
  \langle n\rangle_{\mathrm{meas}} = \frac{\sum_{m}{q_{\mathrm{m}}}
  N_{\mathrm{m}}}{\sum_{m}{N_{\mathrm{m}}}},
\end{displaymath}
where $q_{\mathrm{m}}$ is the calibrated channel number in $p.e.$ and
$N_{\mathrm{m}}$ the number of counts per channel. The number $\langle
n\rangle_{\mathrm{meas}}$ averaged over a series of measurements
amounted to $( 1.55 \pm 0.06)~p.e.$ with a good
reproducibility. Some of the emitted $\gamma$-rays of the cobalt
source undergo Compton scattering and transfer their energy to
electrons which produce Cherenkov light. The maximum electron energy
can be calculated by using $E_e^{\mathrm{max}} = E_\gamma [1 - (1 +
E_\gamma / m_0 c^2)^{-1}] = 890$~keV, where $m_0 c^2$ is the rest mass
of the electron. Electrons with velocities below the Cherenkov
threshold of $\beta_{\mathrm{thr}} = 1/n \approx 0.54$ cannot
contribute to the L.Y.; this limit is equivalent to a minimum electron
energy $E_e^{\mathrm{min}} = 608$~keV. Since the number of emitted
Cherenkov photons increases with the electron energy, a mean electron
energy $\langle E \rangle \approx 800 \pm 40$~keV can be used for
evaluating the detector response. Because significant changes have not
been observed when comparing the cobalt source spectra with the
$\beta$-excited strontium source spectra, the assumed systematic error
in the electron energy of 5\% was confirmed. By dividing the mean
number $\langle n \rangle_{\mathrm{meas}}$ of $p.e.$ by the mean
energy $\langle E \rangle$ a L.Y.\ of 1.9~$p.e.$/MeV was obtained. The
calculated statistical error of the effective L.Y.\ of 4\% is one
order of magnitude larger than the one in the calibration
measurements. This is due to the low count rates using the Cherenkov
radiator \PB\ and could be improved by longer measurements.

In the second part of the study large size $30^2 \times 150$~mm$^3$
crystals were investigated. Fig.~4 shows a typical photoelectron
distribution where the scale is given in numbers of
photoelectrons. Their mean number $\langle n\rangle_{\mathrm{meas}} =
(1.38 \pm 0.05)~p.e.$ corresponds to 1.7~$p.e.$/MeV. The difference
between the result of the sample and the one of the large crystal is
explained by the different light collection efficiency, since this is
a function of the crystal's size, shape and surface finish. The use of
a small sample reduces uncertainties in the light collection process
due to imperfections of the surface, because most of the light
produced inside a sample directly reaches the photocathode.

\subsection{Optimization of the Light Yield}

Since only a small fraction of the produced Cherenkov photons is
detected, the wrapping of the crystals could enhance their
L.Y. However, the Cherenkov light is peaked in the forward direction
with respect to the primary particle's direction and the improvement
is small compared to scintillation counters. The light collection
efficiency for different wrappings was measured with the described
set-up and the cobalt source. Wrapping materials investigated were a
high density, porous, chalk-loaded polyethylene fleece
Tyvek\footnote{{\em Du Pont de Nemours}, Le Grand Saconnex,
Switzerland} in two different thicknesses ($\approx 75~\mu$m and
$\approx 150~\mu$m), two types of PTFE Teflon ($\approx 25~\mu$m and
$\approx 80~\mu$m), a nitrocellulose membrane\footnote{{\em Biometra
biomedizinische Analytik GmbH}, G\"ottingen,Germany} and a
polyvinylidene fluoride named Immobilon-P\footnote{{\em Millipore
GmbH}, Eschborn, Germany} ($\approx140~\mu$m), which is commonly used
as a transfer membrane. The reference L.Y.\ for the comparison was
determined using the unwrapped crystal. The light detected in this
measurement is assumed to originate from internal reflections at the
polished crystals' faces. Then, consecutive layers of the different
reflectors have been added on all five faces. However, it is known
that further layers of material compromise the gain in L.Y.\ due to
the increasing amount of dead material between the crystals, which
makes it possible for shower particles to escape the detector. The
measured photoelectron distributions have been analyzed according to
the method described in the previous section and the results are
presented in Table~1. Using two layers of Teflon tape or one layer of
Immobilon-P gave the highest effective L.Y., confirming earlier
results obtained with a prototype \PB\ calorimeter at the MAMI
Microtron. Both wrapping materials resulted in a 12\% increase
compared to a crystal without wrapping. For this reason the membrane
Immobilon-P was chosen for use in the final detector assembly of the
A4 calorimeter. The membrane has a nominal pore size of 0.45~$\mu$m,
its mechanical strength is barely sufficient. It is hydrophobic, but
it looses its reflectivity when exposed to moisture or coupling
grease.

Since the number of $p.e.$ is strongly affected by the
wavelength-dependent reflectivity of the wrapping material, a
comparative measurement has been carried out with the commercial
double beam spectrophotometer Shimadzu UV-2101 PC. In Fig.~5 the
diffuse reflectance $R$ is shown as a function of the wavelength. The
reflectivity of the Immobilon-P membrane reached 100\% in the visual
area of the spectrum and started decreasing at about 310~nm.

To detect the propagating photons, they have to be transmitted through
an air gap, an optical grease or a glue to the photocathode. The
reflection losses at these boundaries strongly depend on the
difference in reflection indices of the \PB\ crystal and the optical
coupling. To find the coupling with minimum losses in effective L.Y.,
the properties of different optical oils, greases and glues were
studied in the laboratory measurements. The best result has been
obtained with the two-component silicone rubber Elastosil
RT~601\footnote{{\em Wacker-Chemie GmbH}, Burghausen/Obb.,
Germany}. The compound has a viscosity of 5000 \mbox{mPa s}, can be
poured on the crystal, and cures at room temperature during 12
hours. Its refractive index ($ n = 1.41$) is somewhat lower than that
of the entrance window ($n = 1.48$) and significantly lower than that
of \PB\ ($n\approx 1.82$ at 400~nm). Curing a silicon layer of 0.1~mm
thickness in direct contact with the PMT and the crystal not only
provided a good optical coupling but also some adherence of the PMT
and its base.

\section{Monte Carlo Simulations}

\subsection{The Geometrical Set-up and the Detection Method}

For the requirements of the A4 experiment the Cherenkov light
production and detection in \PB\ crystals have been simulated using
the Monte Carlo code GEANT 3.21~[8]. The geometrical set-up used was a
matrix consisting of a $3 \times 3$ array of tapered \PB\ crystals
with $26 \times 26$~mm$^2$ front faces and $30 \times 30$~mm$^2$
readout faces and lengths of 150, 155 and 160~mm, all nine crystals
pointing to the interaction vertex at a distance of about 105~cm. Air
gaps of 300~$\mu$m between adjacent crystals have been
implemented. The number of photons detected by the photon sensor has
to be evaluated from their interactions at the crystals' surface. The
reflection coefficient of the surface finish parameterizes its
reflectivity from perfect smoothness to maximum roughness. A
reasonable number of $90-91\%$ was found by a comparison with
measurements comprising \PB\ crystals at the MAMI electron beam. This
value includes the diffuse reflectance of the Immobilon-P transfer
membrane. The characteristics of 1$\frac{1}{8}$ inch diameter Philips
XP2910 PMTs with borosilicate entrance windows and bi-alkali cathodes
have been used to simulate the photon detection. The internal
transmittancies of the crystals and of the entrance windows have been
measured with the above mentioned spectrophotometer.

To simulate the Cherenkov photon production and detection the GTCKOV
tracking routine of the GEANT code can be used, wherein the photons
are subject to {\em in flight} absorption and medium boundary
action~[8]. This photon transport mechanism requires an evaluation of
the length that the individual photon can travel in the current medium
before each of the possible processes will occur. These numbers are
the different interaction lengths and the minimum among these defines
the step length over which the photon will be transported.  In
addition, the distance from the photon's location to the nearest
boundary has to be calculated and compared with the interaction
lengths. This method usually consumes large computing time, since huge
amounts of produced Cherenkov photons have to be tracked along small
steps to the photocathode.
 
In order to accelerate the Monte Carlo simulation, the characteristics
of the Cherenkov photons have been investigated. First, the number of
generated photons has been determined as a function of the photons'
wavelength~$\lambda$, their angle $\theta$ to the primary particle's
direction and their longitudinal location of production~$z$. This
simulation was done with a full tracking of the Cherenkov photons by
the GTCKOV routine and three-dimensional spectra of produced and
detected photons were obtained. The two-dimensional projection
$\lambda$ vs.\ $z$ of the produced photons inside the $3 \times 3$
array of \PB\ is shown in Fig.~6~(a). The spectrum of the subset of
these photons detected at the photocathode is shown in Fig.~6~(b). The
shape of the latter distribution along the $\lambda$-direction
reflects the transmission of the bulk crystal, the surface properties,
the transmission of the entrance window and finally the quantum
efficiency of the PMT. The distribution exhibits a steep rise at
270~nm, followed by a maximum at 330~nm, and a slower decrease up to a
wavelength of about 600~nm. The shape of the distribution along the
$z$-direction corresponds to the longitudinal profile of the Cherenkov
photon shower. To obtain the detection probability $p$ the
three-dimensional spectrum of detected photons was divided by the
spectrum of produced photons. The resulting probability distribution 
is shown in Fig.~7. The full tabulated probability distribution
$p(\lambda,\theta,z)$ has been implemented as a look-up table in the
simulation code, so that for subsequent simulations the probability
$p$ of all produced Cherenkov photons could be compared with a
randomly generated number. If $p$ was larger than the random number,
it was assumed that the photon could be detected, otherwise it was
supposed to be absorbed. This method avoids the tracking of photons,
and this reduces the computing time by a factor of fifty, which allows
to simulate the detector response and more complex geometries in
reasonable time scales.

\subsection{Simulated Properties of the Cherenkov Light}

The half-angle $\theta_{\mathrm{Ch}}=\arccos(1/n\beta)$ of the
Cherenkov cone is a characteristic observable for a particle with
velocity $v = \beta c$ in a medium with the index of refraction
$n$. Depending on the lateral development of the electromagnetic
shower particles an angular distribution of the emitted Cherenkov
photons as shown in Fig.~8 was evaluated. The maximum of the
distribution at 56$^{\circ}$ is clearly pronounced, which corresponds
to the half-angle $\theta_{\mathrm{Ch}} = 57^{\circ}$ of relativistic
leptons, but tails of the distribution reach 0 and 180 degrees. The
calculated number of radiating leptons per event was $\langle
n\rangle_{\mathrm{lep}} \approx 219$~lep./GeV. The distribution of the
number of photons produced per centimeter path length extended from
only few photons up to a limit of $dN/dx \approx 900$~photons/cm. The
strong rise to the limit is due to the low rest mass of the electrons
and positrons. The value of the obtained limit is in accordance with
the theoretical number, which can be calculated by the following
equation~[10]:
\begin{displaymath} 
  \frac{d^{2}N}{dxd\lambda}=\frac{2 \pi \alpha
  z^{2}}{\lambda^{2}}\left(1- \frac{1}{\beta^{2} n^{2}(\lambda)}
  \right)
\end{displaymath} 

The lateral distribution of the Cherenkov photons was found to be
narrower than the distribution of the energy deposition in the
electromagnetic shower.  This is caused by the decreasing fraction of
energy which is carried by the leptons with increasing depth inside
the crystals~[10] and due to the larger fraction of leptons with
increasing lateral extension that have energies below the Cherenkov
threshold of 608~keV. This results in an apparent Moli{\`e}re radius
$R_{\mathrm{M}} \approx 1.8$~cm which is smaller than the nominal
radius $R_{\mathrm{M}} = 2.2$~cm (see e.\,g.\ [2]).

\subsection{Light Yield and Detector Response}

Cherenkov radiators are characterized by their effective L.Y.\ and
their intrinsic non-linearity. The best energy resolution will be
achieved when the L.Y.\ is large and proportional to the energy of the
primary particle. Since $\langle n\rangle_{\mathrm{lep}}$ is small,
its fluctuations $\delta n_{ \mathrm{lep}} = 6$\% at 734~MeV contribute
considerably to the energy resolution of the crystals. The total
number of produced Cherenkov photons amounted to $n_{\mathrm{Ch}}
\approx 20,000$~photons at 855~MeV, which is equivalent to 23.4
photons/MeV. This number, however, is of low significance, because the
photons are subject to a multitude of processes reducing the number by
about 90\%. For that reason it is more interesting to determine the
number of $p.e.$ detected by a photon sensor of given type and
sensitivity. This effective L.Y.\ is a function of $n_{\mathrm{Ch}}$
and depends on the light collection efficiency, the quantum efficiency
and the efficiency of the photoelectron collection. In a $3 \times 3$
array the simulated L.Y.\ at 855~MeV was $\langle
n\rangle_{\mathrm{sim}} = (2163 \pm 2)~p.e.$\ which corresponds to
an effective L.Y.\ of $2.53~p.e.$/MeV. By using a set-up consisting of
only a single crystal a L.Y.\ of 2.10~$p.e.$/MeV was calculated.  This
value was simulated at energies above 10~MeV and could be compared
with the low energy measurements, but one has to be aware that the
simulation does not include small imperfections in the light and
photoelectron collection.  In contrast, the real crystals exhibited
minor defects in the bulk material and on the surface. In addition,
the detector response is degraded at such low electron
energies. Together both facts presumably explain the small difference
between the simulated and measured L.Y. 

Calculations using different electron energies provided a measure of
the detector response and its differential non-linearity, which is
defined by the variation in the L.Y.\ as a function of the energy of
primary particle. At energies between 10 and 1000~MeV the L.Y.\
increased monotonically between 2.52 and 2.54~$p.e.$/MeV. The
projection of the detection probability distribution on the axis of
the photon's longitudinal location of production~$p(z)$ is interesting
in terms of the non-linearity of the L.Y. As can be inferred from
Fig.~7., its slope in $z$-direction is 0.12\% per cm at the position
of the shower maximum at 5~cm inside the crystal ($z \approx$ 110~cm).

\section{Summary}

The L.Y.\ of the Cherenkov radiator \PB\ has been studied, because the
parity violation experiment at MAMI requires a good energy resolution
of its electromagnetic calorimeter and a high L.Y.\ of their
crystals. It is shown that a HPMT can be used to obtain the effective
L.Y. Measurements with a low energy $\gamma$-source as well as with a
$\beta$-source revealed a number of $1.7-1.9~p.e.$/MeV with 4\%
statistical and 5\% systematic error. The L.Y.\ was increased by using
appropriate wrappings on the crystal and a well adapted coupling to
the photon sensor. The latter results confirmed that the Immobilon-P
membrane is the best wrapping material to be used with Cherenkov
crystals.

Monte Carlo simulations have been performed in order to evaluate the
L.Y.\ of an electromagnetic calorimeter consisting of $3 \times 3$
\PB\ crystals. By establishing a method to accelerate the GEANT code a
computing time reduction by a factor of fifty was
achieved. Nevertheless, a simulation with full Cherenkov photon
tracking was still needed to obtain the detection probability
distribution. This approach was used to simulate various experimental
implications and geometrical effects. An effective L.Y.\ of
$2.1~p.e.$/MeV for a single \PB\ crystal was calculated, which agrees
fairly well with the HPMT measurements, taking small imperfections of
the crystal and a lower detector response at very low energies into
account. The L.Y.\ is very high compared to other Cherenkov radiators
and is needed to obtain the good energy resolution required by the
parity violation experiment. Besides, the simulations predicted a very
high linearity of the response in the energy range between 10 and
1000~MeV with a variation smaller than 1\%. This promising result
confirms that \PB\ is very well suited for its use as a calorimeter
material, since larger non-linearities would have degraded the energy
resolution.

\begin{ack}
  The authors would like to thank J. Garcia from University of Valencia
  for his help during parts of the HPMT measurements.
\end{ack}

\newpage
\begin{figure}[htbp]
   \begin{center}
     \subfigure[Set-up during measurements with a cobalt
        $\gamma$-source. The HPMT and the coincidence counter
        triggered the readout.]{\psfig{figure=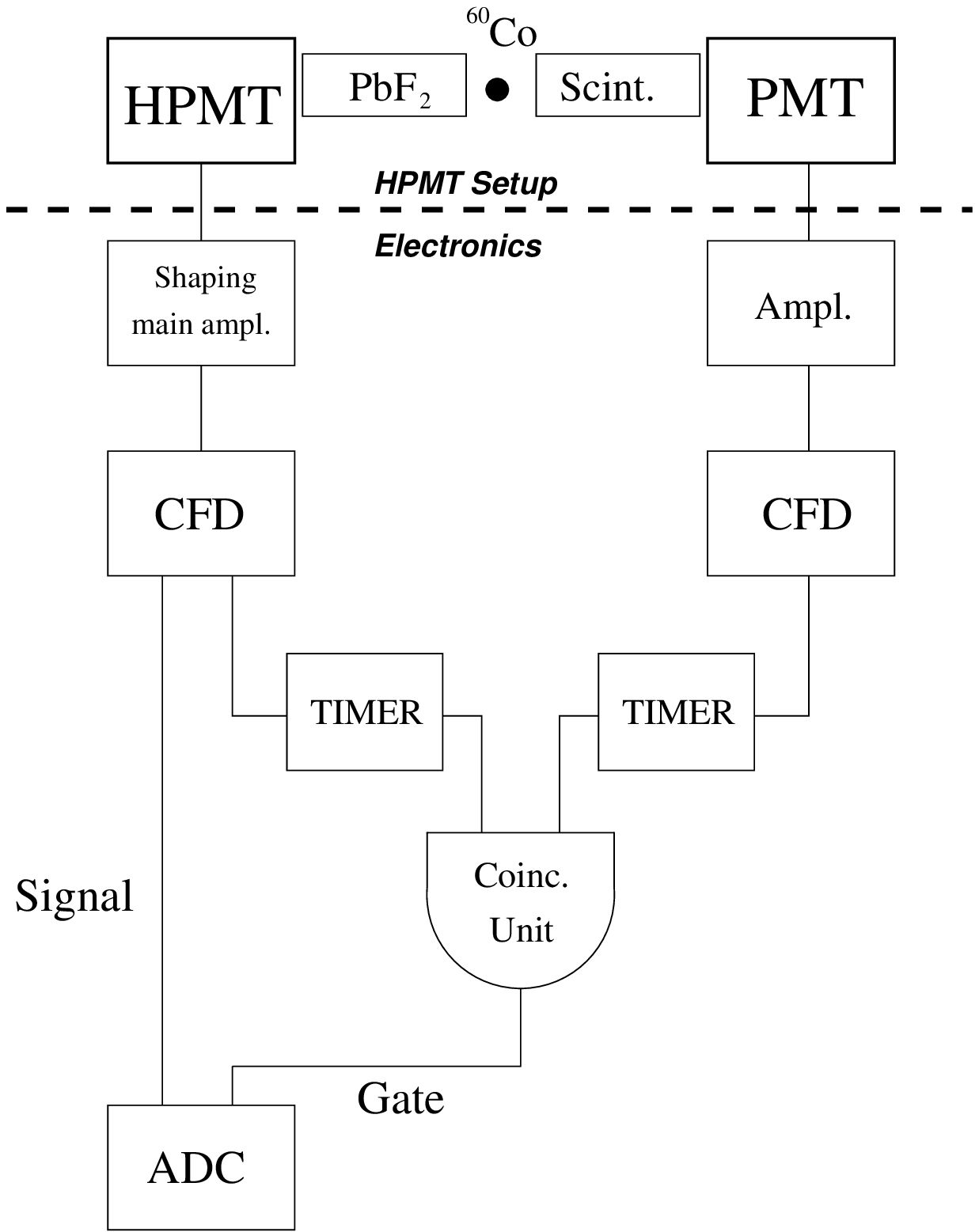,
        width=0.45\textwidth}}
     \subfigure[Set-up during measurements with a strontium
        $\beta$-source. The two coincidence counters triggered the
        readout.]{\psfig{figure=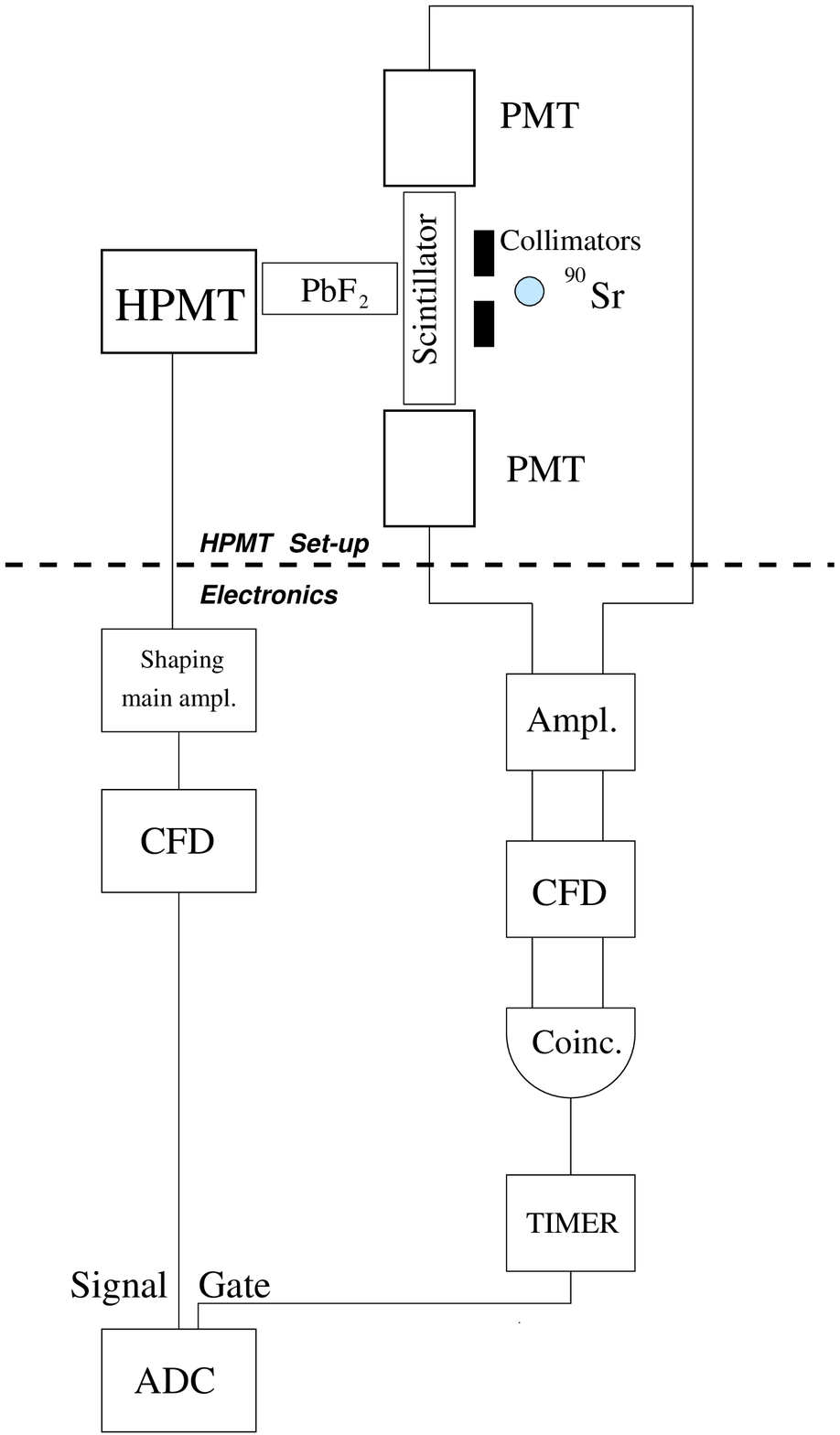,
        width=0.45\textwidth}}
   \end{center}
   \caption{HPMT detector set-up and block diagrammes of the readout
     	electronic components for the light yield measurements. Standard
        CAMAC electronics was used, CFD abbreviates {\em Constant 
     	Fraction Discriminator}, ADC {\em Analog-to-Digital Converter.}}
\end{figure}

\begin{figure}[htbp]
   \begin{center}
     	\mbox{\psfig{figure=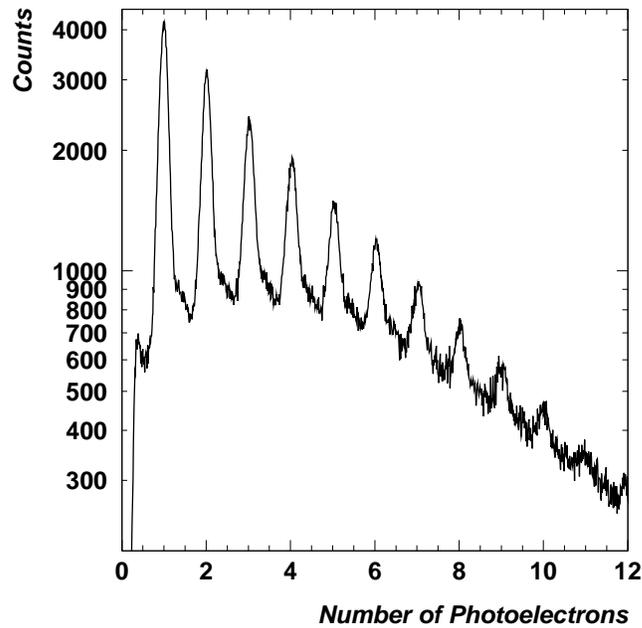,
	width= 0.6 \textwidth}}
   \end{center}
   \caption{Photoelectron distribution of highly attenuated LED pulses
     	that have been measured with the HPMT. The distribution is
     	composed of Gaussian shaped peaks and a continuum. Note the
     	large peak-to-valley ratio of the first photoelectron peak.}
\end{figure}

\begin{figure}[htbp]
   \begin{center}
     	\mbox{\psfig{figure=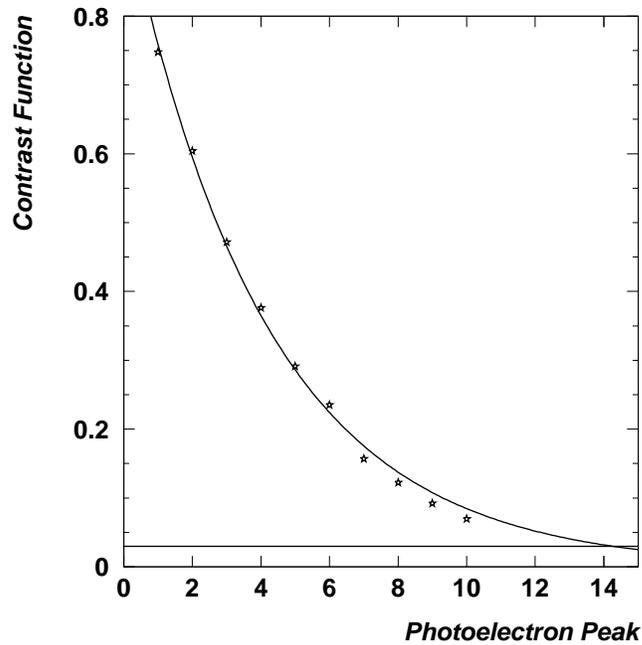,
	width= 0.6 \textwidth}}
   \end{center}
   \caption{The HPMT contrast function $f$ which was calculated from the
     	photoelectron peaks of the LED pulse measurements. The
     	straight solid line at 0.03, which is commonly defined as the
     	limit of peak resolution, crosses the exponentially fitted
     	data points at 14 $p.e.$}
\end{figure}

\begin{figure}[htbp]
   \begin{center}
     	\mbox{\psfig{figure=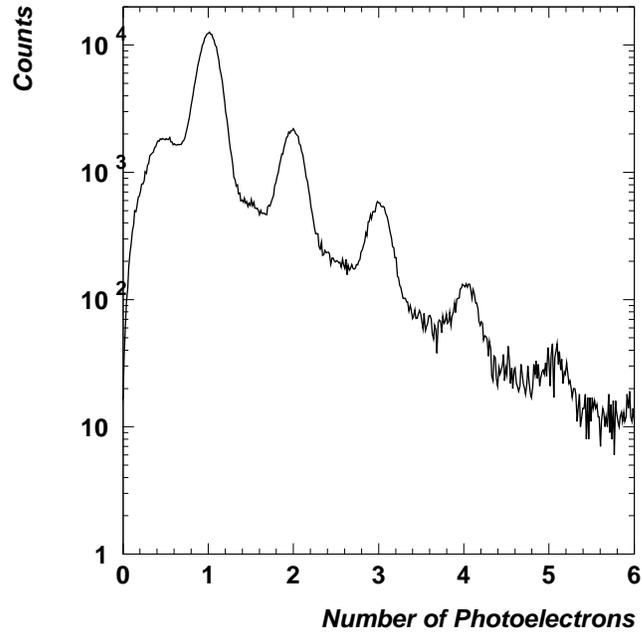,
	width= 0.6 \textwidth}}
   \end{center}
   \caption{A typical photoelectron distribution of a \PB\ crystal
     	from the series of measurements with the \nuc{60}{Co}
     	$\gamma$-source. The distribution corresponds to the mean number
     	of photoelectrons $\langle n\rangle_{\mathrm{meas}} = (1.38 \pm
     	0.05)~p.e.$ from which a L.Y.\ of 1.7 $p.e.$/MeV was derived.}
\end{figure}

\begin{figure}
  \begin{center}
    	\mbox{\psfig{figure=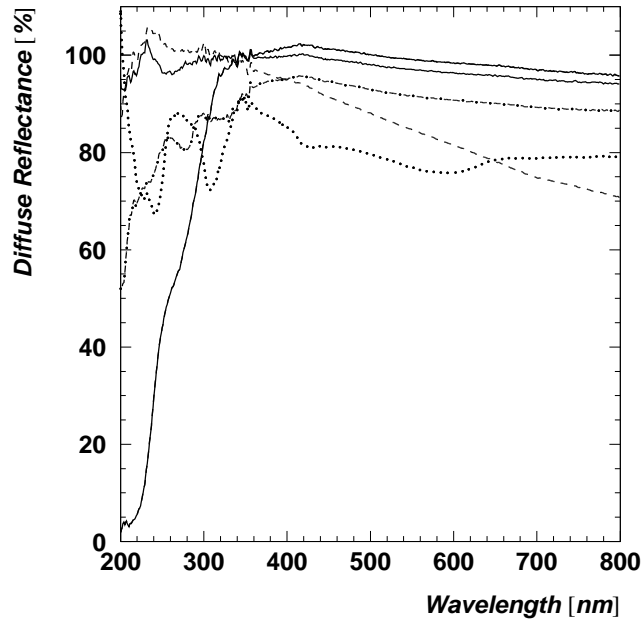,
	width= 0.6 \textwidth}}
  \end{center}
  \caption{Diffuse reflectance of wrapping materials. The different
    	types of material are encoded as follows: solid line =
    	Immobilon-P ({\em Millipore}); dot-dashed = Tyvek ({\em Du
    	Pont}); narrow dots = office paper; dashed = Teflon; wide 
    	dots = nitrocellulose membrane ({\em Biometra}).}
\end{figure}

\begin{figure}[htbp]
   \begin{center}
     \subfigure[Distribution of photons produced inside the
        crystal.]{\psfig{figure=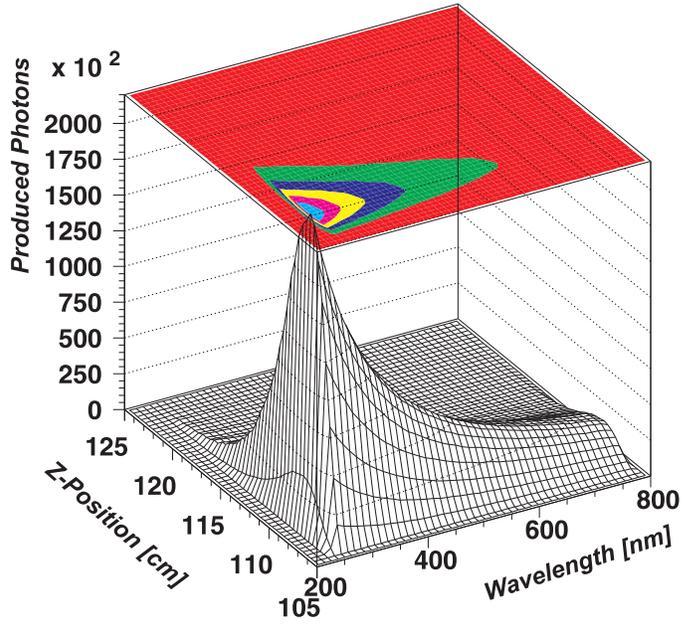, 
	height= 0.36 \textheight}}\\
     \subfigure[Distribution of photons detected in the
        photomultiplier.]{\psfig{figure=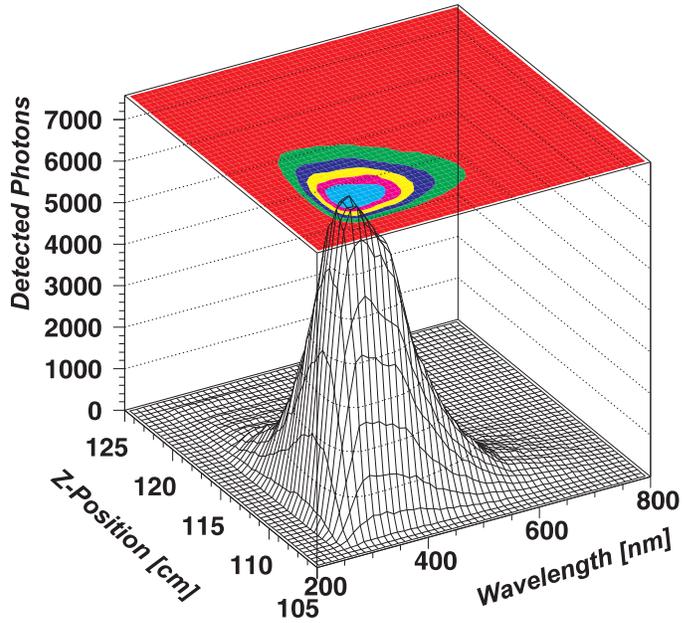,
        height = 0.36 \textheight}}
   \end{center}
   \caption{Two-dimensional spectra of the Cherenkov photon simulation.
        The location $z$ of the photons' production along the longitudinal
        axis is plotted versus their wavelength $\lambda$.}
\end{figure}

\begin{figure}[htbp]
   \begin{center}
     	\mbox{\psfig{figure=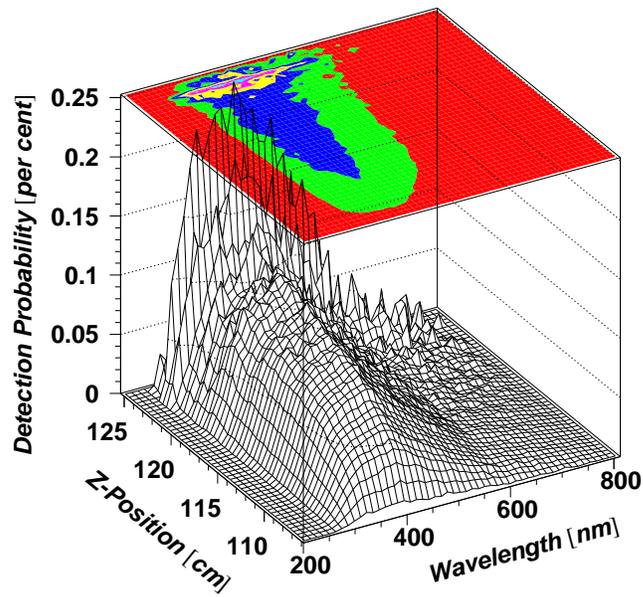,
	width= 0.6 \textwidth}}
   \end{center}
   \caption{Detection probability of the produced Cherenkov photons as
     	a function of the location of the photon's production and
     	their wavelengths. At the position of the maximum of the 
	electromagnetic shower at 5~cm inside the crystal ($z \approx$
     	110~cm) the slope along the $z$-direction is 0.12\% per cm.}
\end{figure}

\begin{figure}[htbp]
   \begin{center}
     	\mbox{\psfig{figure=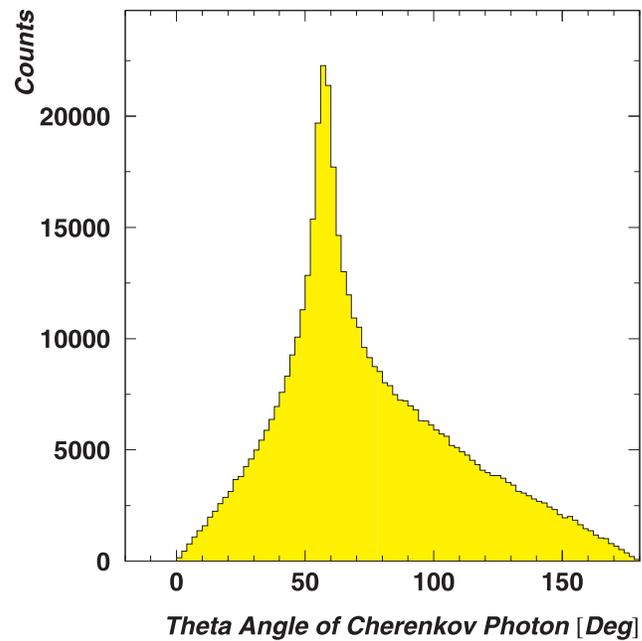, 
	width= 0.6 \textwidth}}
   \end{center}
   \caption{Simulated spectrum of the Cherenkov photon's angle
     	$\theta$ with respect to the direction of the primary
     	particle. The peak at 56$^\circ$ agrees well with the
     	half-angle $\theta_{\mathrm{Ch}} \approx 57^\circ$
     	of the Cherenkov cone for relativistic particles.}
\end{figure}

\newpage
\begin{table}[htbp]
  \caption{Effective light yield (L.Y.) of a \PB\ crystal wrapped in
	several layers of different reflective materials. The
	thickness of the wrapping defines the dead material between
	adjacent crystals.}
  \begin{tabular}{llll} 
   \\
   \hline 
   Wrapping Material & No. of Layers & Thickness [$\mu$m]& L.Y. [\%]\\
   \hline
   \hline	
   Immobilon-P          & 1             & 140           & 112.1 \\ 
   Teflon Tape          & 2             & 160           & 112.1 \\
   Teflon + Al          & $4+1$         & $100+35$      & 111.5 \\
   Teflon               & 3             &  75           & 110.5 \\
   Teflon Tape          & 1             &  80           & 109.7 \\
   Tyvek                & 1             & 140           & 108.6 \\
   Teflon + Al          & $3+1$         & $75+35$       & 108.0 \\
   Tyvek                & 1             &  80           & 106.9 \\
   Teflon               & 2             &  50           & 106.5 \\
   Office Paper         & 1             & 100           & 105.8 \\ 
   Teflon               & 1             &  25           & 102.7 \\
   \hline
   Unwrapped            &               &               & 100.0 \\
   \hline
  \end{tabular}
\end{table}

\end{document}